# Prediction of amino acid side chain conformation using a deep neural network

Ke Liu [1], Xiangyan Sun[3], Jun Ma[3], Zhenyu Zhou[3], Qilin Dong[4], Shengwen Peng[3], Junqiu Wu[3], Suocheng Tan[3], Günter Blobel[2], and Jie Fan[1],

Corresponding author details:

Jie Fan,

Accutar Biotechnology,

760 Parkside Ave, Room 213

Brooklyn,

NY 11226, USA.

jiefan@accutarbio.com

1. Accutar Biotechnology,

   760 parkside Ave, Room 213, Brooklyn, NY 11226, USA.

2. Laboratory of Cell Biology, Howard Hughes Medical Institute, The Rockefeller

   University, New York, New York 10065, USA

3. Accutar Biotechnology (Shanghai)

   Room 307, No. 6 Building, 898 Halei Rd, Shanghai, China

4. Fudan University

   220 Handan Rd, Shanghai, China

**Summary:**

A deep neural network based architecture was constructed to predict amino acid side chain conformation with unprecedented accuracy.

Amino acid side chain conformation prediction is essential for protein homology modeling and protein design. Current widely-adopted methods use physics-based energy functions to evaluate side chain conformation. Here, using a deep neural network architecture without physics-based assumptions, we have demonstrated that side chain conformation prediction accuracy can be improved by more than 25%, especially for aromatic residues compared with current standard methods. More strikingly, the prediction method presented here is robust enough to identify individual conformational outliers from high resolution structures in a protein data bank without providing its structural factors. We envisage that our amino acid side chain predictor could be used as a quality check step for future protein structure model validation and many other potential applications such as side chain assignment in Cryo-electron microscopy, crystallography model auto-building, protein folding and small molecule ligand docking.

(word count: 130/150 words)

**Introduction**

Prediction of amino acid side chain conformations on a given peptide backbone is essential for protein homology modeling, protein-protein docking(*1*), protein *ab initio* folding (*2*), and small molecule drug docking and design(*3, 4*). Over the past 20 years, many computational methods have been developed to solve the fundamental problem of side chain prediction(*5, 6*). Historically, side chain prediction involves two steps. First, a side-chain conformation library (rotamer library) is constructed based on statistical clustering of observed side chain conformations in the protein data bank (PDB), allowing the side chain being predicted to sample in this artificially constructed search space(*7*). Second, a physics-based scoring function is used to evaluate the likelihood of the sampled conformations(*8-11*). Of the prediction methods currently available, Side Chain With Rotamer Library 4 (SCWRL4)(*6, 12*) is the most widely-used method because it is accurate and fast.

We believe that the side chain prediction problem has been largely overlooked, in part, due to the use of relatively less-stringent evaluation criteria. Using current standards, a prediction is considered correct if the predicted side chain position has a Chi angle within 40 degrees of the X-ray positions(*13*). The reported performance for the current standard method, SCWRL4, is ~90% according to this criterion(*6*). Additionally, the SCWRL4 method predicts side chain conformations without providing variances of the estimate, which limits the justification of the method itself. More importantly, aromatic residues, such as tyrosine and tryptophan, are especially sensitive to these types of Chi-angle based errors. In addition, the SCWRL4 algorithm determines disulfide bonds before other types of bonds(*6*), which lacks biological foundations and will potentially introduce errors.

Thanks to the structural genomic initiative, the deposit number in the PDB database has seen explosive growth in the past decade with over 100,000 protein structures now available(*14*). This has been accompanied by the development of more transformative statistical analysis tools such as deep neural networks(*15, 16*), which have been shown to surpass human performance in multiple tasks from object recognition to strategic board games such as Go(*17*).

We decided to tackle this old side chain prediction problem using a more data-driven approach. In this paper, we outline the development of a deep neural network architecture for side chain conformation prediction. We first classified each amino acid side chain into a backbone-independent rotamer library. By further modeling amino acids side chains with 3-Dimensional (3D) images, we were able to use a deep neural network to predict the likelihood for targeting amino acids adopting each rotamer. The most likely rotamer ranked by our convolutional neural network (CNN) architecture was the output for the prediction. Using this approach, we were able to improve side chain prediction accuracy by more than 25% according to an unbiased Root Mean Square Deviation (RMSD) calculation. More importantly, when we modeled the distribution of the prediction score of a large training set, our platform not only provided a favorable rotamer for a side chain in a given environment, but it also provided information on how likely the side chain adopts a certain rotamer. This statistical property of the predictive score enabled us to perform a pan-PDB database side chain quality evaluation without supplying structure factor information. As a result, we were able to identify thousands of conformational outliers for each amino acid type in the database, including clashes, mis-assigned conformers or residues that lack electron density. Many of the conformational outliers have been independently confirmed by real space validation methods including real-space R-value Z-score (RSRZ) methods (*18*).

**Results and discussion**

**Construction of the amino acid rotamer library**

Historically, the side chain conformation prediction problem has relied on efficient clustering of available side chain conformations. As a result, the side chain prediction problem has been reduced to a side chain subclass assignment problem. In practice, an ideal rotamer library should satisfy the following requirements: the number of the rotamer should be kept as small as possible, in order to enable efficient searching of side chain conformations; and the average RMSD between the true conformations and their most similar rotamers in the library should be as small as possible, in order to ensure the accuracy of predicting side chain conformations. Current popular methods include the use of back-bone independent(*19*) and back-bone dependent rotamer libraries(*7, 13*).

Amino acids have 1 to 4 Chi angles, depending on the lengths of the respective side chains. Accordingly, the SCWRL4 side chain rotamer library is constructed in a hierarchical manner along the multiple Chi angles of each side chain(*13*). As a result, in the SCWRL4 side chain rotamer library, Arg has 81 subclasses and Phe has 27 subclasses. We reasoned such a hierarchical classification method may be spatially too sparse to cover enough conformational space. Instead, we adopted a flat structure to classify the side chain conformations of each amino acid based on the geometrical differences among the side chain conformations using a k-means clustering algorithm(*20*). Details on the Accutar rotamer library is provided in **Supplementary Table 1**.

Protein conformation can be encoded by both backbone information and side chain conformation. Since the backbone conformation encoded in 3D image format was going to be used in our CNN model, we decided to construct the amino acid side chain rotamer library in a backbone-independent fashion, which also reduced the number of side chain rotamers. In fact, using this strategy, we were able to classify the side chains into fewer classes, which covered conformational space more efficiently as shown by the theoretic limit cumulative distribution functions (CDF) plot (**Fig.1**). In this analysis, the theoretic limit CDF function measures the probability of a theoretical model deviates from its genuine structure at a certain RMSD cutoff. The CDF was defined as follows:

$$F_X(x) = P(X_{deviation}(RMSD\ measured\ in\ \text{Å}) \leq x)$$

This calculation was based on the assumption that the side chain conformations of all amino acids in a protein have been fully represented by the rotamer nearest to the genuine conformation. Hence, the deviation (measured by RMSD) between the genuine structure and models, represented by different rotamer libraries including the SCWRL4 rotamer library, Duke rotamer library or Accutar rotamer library, could be measured and used to calculate the CDF functions. Under this assumption, an ideal classification method should produce a lower theoretical limit RMSD using a relatively low number of classes. As shown in **Fig. 1**, across all amino acid types, using a RMSD cutoff of 0.5 Å, the Accutar rotamer Library (colored red in **Fig. 1**) covered more conformational space than the current standard backbone-dependent rotamer library, the SCWRL4 Rotamer Library by ~20% (colored blue in **Fig.1**, right panel) and outperforms the backbone-independent Duke Rotamer Library (colored blue in **Fig. 1**, left panel) by ~25%. The RMSD values measuring the deviation from genuine structure for each amino acid of Accutar rotamer library versus those derived from the SCWRL4 rotamer library are provided in **fig. S1**.

**Construction of neural network architecture**

To model amino acid side chain conformation feasibility with 3D images, side chains were encoded by 23 atom types which can be considered as 23-color channels for the image. The 23 Atom types are chosen based on basic chemical properties and chemical environment of each atoms. The detailed parametrization procedure is explained in the **Methods** section. Through this parametrization procedure, the side chain conformation prediction problem could be considered as an image processing problem for which the CNN method has been successfully integrated previously(*21, 22*).

We used a 3D CNN architecture implemented with the Microsoft Cognitive Toolkit (CNTK)(*23*) to model the protein side chain conformation from its environment (**Fig. 2a**). This 9-layer network accepts graphic input of an amino acid adopting certain rotamer with its environment, and it outputs a probability score of different potential rotamers.

Every input amino acid was aligned by their Cα, amine and carboxyl group so that the amino acid to be predicted and its neighboring environment were firstly quantized into a 3D voxel grid(*24*) representing the position and interaction of all related atoms. The voxel grid was then fed through several 3D convolutional and pooling layers to predict a feasibility score for each conformation. The modeled feasibility score was then trained over a large protein structure database so that different conformations could be compared to predict the most favorable conformation of an amino acid given its environment.

To understand how the CNN model represents and learns useful atomic interaction features, we analyzed the trained CNN model by visualizing its convolutional layer filters(*23*). The input patches, which maximally activates a filter in the first convolution layer, are shown (**Fig. 2b**). Each group of five patches in one column in the figure corresponds to a single filter in the first convolution layer. The red cube designates the input region. As can be seen, the neural network was able to capture many interesting and useful features, such as disulfide bonds (left panel of **Fig. 2b**), benzene rings (middle panel of **Fig. 2b**), and electrostatic interaction (right panel of **Fig. 2b**). The fact that the CNN model learned these useful chemical moieties without any prior chemistry domain knowledge partially explains how the CNN model learns the concrete image models of amino acid side chains.

**Internal ranking model performance of CNN architecture**

The CNN architecture centered on a ranking model-based training algorithm (the detailed ranking algorithm is provided in the **Methods**), because, for every querying residue with an amino acid type specified, the CNN needed to rank the likelihood of all possible rotamers in that specific position. The internal ranking model performance with respect to different amino acid types are provided in **fig. S2**. We noted, the ranking model had a precision rate of ~60% for top picks for most amino acid types including aromatic residues, whereas the performances were relatively modest in charged amino acids. More specifically, the precision rates occurred in the range of 30−40% for top picks for Lys, Glu and Gln which suggested there is room to improve on these residues in the future.

**Leave-one-out (LOO) side chain prediction test to evaluate the predictor**

Traditionally, the performances of different protein side chain conformation prediction programs are hard to compare due to different judgement criteria or different testing sets used. In order to evaluate our CNN method head-to-head against the current popular SCWRL4 method, we adopted a more unbiased leave-one-out (LOO) test, using the same 379 PDB testing datasets from the original SCWRL4 paper. To avoid using evaluation data in the training procedure, the structures with a sequence similarity of 70% of any testing structures were excluded from CNN training sets. Using this approach, the two methods were allowed to run a sequential prediction for every individual amino acid along the protein sequence with all other residues conformations given for each test. After the LOO test was allowed to run through all the structures in the testing set and instead of using the Chi-angle criteria used previously, we decided to use a more unbiased RMSD criteria to evaluate the deviations between the predicted model and experimentally determined model (set as ground truth), which allowed us to compare the relative performance of the SCWRL4 method and the CNN method. Overall, we noticed that the CNN method outperformed the SCWRL4 method in all 20 amino acid subtypes in RMSD values (**Fig. 3**). Using the RMSD value of 0.5 Å as a cut-off (i.e. by only comparing the accuracy rate of predicted side chain conformation deviates from the observed side chain conformation by a per-atom distance of within 0.5 Å RMSD range), the CNN method on average showed ~25% higher accuracy rate than that of SCWRL4 method. More striking performance improvements of ~40% were observed in aromatic residues and long side-chain residues (**fig. S3**). To our knowledge, performance improvement of this scale is unprecedented since the side chain prediction problem first surfaced 20 years ago.

**LOO-score as a structure model quality indicator**

We then aimed to determine whether the CNN-based amino acid side chain predictor has other applications in structural biology. First, we looked at the distribution of average LOO score of all PDB structures. The LOO score assumed a unimodal distribution skewed to the right (**Fig. 4a**). To the far-left end of the x-axis, the structures with poor LOO scores are enriched with nuclear magnetic resonance (NMR) structure models; electron microscopy (EM) and Cryo-electron microscopy (EM) models occur next in the higher LOO score region, followed by X-ray structure models with higher resolution more or less localized to the right side of the figure (**Fig. 4b**). (The LOO scores for different groups of structures were plotted in **fig. S4**.) This distribution pattern could be explained by the lack of efficient real space side chain conformation refinement steps in NMR and EM structure model building. The LOO score has an excellent linear relationship with resolution of structure models with an $R^2$ of ~0.5 for sample size of ~50,000 models (**Fig. 4c**). We then aimed to determine whether the LOO scores for individual side chains deposited in PDB database could be used as a side chain model quality metric for the following reasons: First, at present, side chain model quality can only be verified by Ramachandra statistics and by checking the deviations between the model and electron density map in real space; Second, we observed a strong correlation between the model LOO score and model quality. We therefore calculated the individual side chain LOO scores of all PDB structures using deposited conformations. These scores were ranked by how much the observed LOO score deviates from the mean value of the LOO score calculated from CNN training process, we were able to pick up thousands of LOO score outliers from published structures (**Fig. 5**). A detailed list and respective maps for the top amino acid outliers with an associated

LOO score beyond 3 standard deviation ranges of the average LOO score of its kind, are provided in the **Supplementary Table 2**. By systematically examining the top outliers for each amino acid type, we were surprised to see that ~60% of the outliers fell into the following three categories:

Steric clashes which account for ~4% of outliers detected (shown in blue in **Fig. 5a**), residues with mis-assigned conformers as independently confirmed by RSRZ outlier analysis(*18, 25*) which accounted for ~8% of the outliers (shown in green), and mis-assigned conformers not identified by RSRZ outliers which accounted for ~48% (shown in cyan). Steric clash errors could be easily verified by checking the model itself; however, it was difficult to distinguish between Ramachandran and rotamer outliers. We therefore decided to further explore the second type of error (possible mis-assignment of side chain conformations). Some examples of electron density maps were shown in **Fig. 5b.** In these figures, predicted models are shown in brown and deposited models are shown in green. The 2Fo-Fc maps contoured at 1.0 sigma are shown in blue and the Fo-Fc maps contoured at 3.0 sigma shown in red/green. In all cases, the predicted side chain rotamers pointed to the positive density region (in green) whereas original rotamers deposited in the database were in the negative electron density area (in red). The pie charts summarizing LOO outliers for each amino acid were provided in **fig. S5**.

Here, we demonstrate for the first time the use of deep learning to accurately predict amino acid side-chain conformations. The LOO statistics described here allowed us to systematically compare our method with the current standard SCWRL4 method. In this large-scale test, our CNN platform improved prediction accuracy by over 25% across all amino acid types. This ability to identify conformational outliers deposited in the PDB without supplying structure factors suggests its potential applications in multiple fields from structural model validation, structural model auto-building in crystallography and Cryo-EM through to side-chain flexible mode small molecule docking.

## Methods

### Graphics

All structural figures were prepared by Orbital (Accutar Biotech).

### Atom Type

Atom type is a unique index assigned to each atom in a polymer, including both atoms of amino acids and hetero atoms. The mapping table between atoms in a polymer and the atom types is provided in **Table 1**. In general, atoms of different elements will have different atom type indices, while atoms of the same element may also have different atom type indices if these atoms are chemically different or in a different environment. Atom types allow abstraction of atoms of different amino types.

### Datasets

All available PDB data files were used to derive atom types and the rotamer library. The evaluation dataset was the same as used by SCWRL4. The training dataset was generated by using all public structures derived using X-ray crystallography from the Research Collaboratory for Structural Bioinformatics (RCSB) PDB(*21, 26*), excluding those with a resolution above 1.7Å, those with missing atoms or having clashed atoms, and those with chains similar to one in the evaluation dataset. There was a total of 12809 PDB files and ~3,840,000 amino acids in the training dataset, and 379 PDB files and ~72,000 amino acids in the evaluation dataset.

**Data preparation**

Only structures obtained using X-ray diffraction were kept. Symmetry mates were added to the original protein structure prior to training and evaluation to restore the original crystal structure environment.

**Input quantization**

Every input conformation was represented as a grid of 20*20*20 voxels, each voxel representing a 1 Å$^3$ volume. Each atom in an amino acid and related environment is represented as a smoothly interpolated sphere in the grid, using the soft-bin fill algorithm (**fig. S6**). Each of the 23 atom types form a channel in the input feature map. Atoms of the side chain conformation to be predicted and of its environment are extracted into separated channels to be able to distinguish them. Therefore, a total of 46 input channels are used.

The soft-bin grid fill algorithm takes an input atom and fills the voxel grid region the atom occupies. The occupation ratio is obtained by treating the atom as a 1x1x1 cube and calculating the intersection volume between the cube and a voxel. The occupation ratio is further normalized to make sure all occupation ratio of an atom sums up to one.

**Negative sampling**

The original conformation of each amino acid in the training dataset was set to be ground truth. In order to obtain negative samples, we use a hybrid global and local sampling approach, unlike SCWRL4 which only uses a global conformer library. As with previous approaches, we first obtained a conformer library by aggregating and clustering amino acid conformations from all available protein structure data. Using this library, we are able to sample many different conformations for a given ground truth. However, since the conformer library is globally averaged, and due to the fact the potential number of conformers is very large (up to 4 dihedral angles), globally clustered conformer library is insufficient in some cases. To overcome this issue, we additionally use an algorithm to perturb the conformation of amino acids and obtain localized negative conformations. The perturbation algorithm is shown in **fig. S7**. The perturbation algorithm starts with a perturbation angle predefined by the type of the amino acid. Then it iteratively processes each dihedral angle. For each dihedral angle, it generates two samples by rotating the dihedral angle by the perturbation angle back and forth. A decay is applied after each dihedral. This procedure gives more flexibility to dihedral angles in the far end than dihedrals near the backbone.

**Training algorithm**

All training data was organized as $<a, b>$ pairs such that the conformation $a$ should be ranked better than conformation $b$. Several types of ranking pairs were extracted for training:

1. Ground truth rotamer (the closest rotamer in the rotamer library to the ground truth) was ranked better than all other rotamers in the rotamer library.

2. Ground truth was ranked better than the most similar rotamer in the rotamer library.

3. Ground truth was ranked better than all locally perturbed conformations.

During ranking pair generation, if the RMSD between the two conformations was lower than predefined threshold, the pair was thought to be ambiguous and discarded from the training dataset. This may happen, for example, when the ground truth is very similar to the ground truth rotamer, in which it is impossible to determine which one is better.

We used Microsoft's CNTK toolkit for training the neural network. The neural network takes input a voxel grid of quantized amino acid environment and approximates a piecewise ranking score. The 20*20*20 voxel is fed through a 3*3*3 convolutional layer and a 5*5*5 convolutional layer, with a 2*2*2 max pool subsampling. Then another 3*3*3 and 5*5*5 convolutional layers are applied. Finally, a global average pooling layer is used to aggregate information from the entire grid and several fully connected layers are applied subsequently to project the output to a scalar score. Rectified Linear Unit (ReLU) non-linearity is used throughout the process except the output layer, where a sigmoid non-linearity is used to map the output to probability of range (0, 1).

During training, the scores of the ranking pair $a$ and $b$ are calculated and compared. The training loss is defined to favor correct pairwise ranking predictions.

**Inference algorithm**

During inference, the correct conformation of an amino acid given its ground truth environment needs to be predicted. This was carried out using a two-phase algorithm. First, all global rotamers for the amino acid were sampled with their conformational score predicted. The conformer with the best score was kept as the most probable conformer. Second, the conformer is further optimized through an iterative fine tuning process. In each iteration, all possible perturbation with angle $\alpha$ of current conformer is enumerated, then evaluated. The one with highest score is kept for next iteration. The angle $\alpha$ is divided by 3 after each iteration, so that each conformation enumerated is not the same with each other, and all conformations are uniformly covering the dihedral combinations similar to the initial one. The details of fine tuning algorithm are shown in **fig. S8**.

The fine tuning algorithm starts with an amino acid and a predefined maximum depth. It generates samples by enumerating all combination of chi angle rotations with a certain angle interval. A decay rate is applied to the perturbation angle as the Perturb algorithm.

**Supplementary materials** are available at http://open-file.accutarbio.com/sidechain_supp_for_download_v1.zip

**Acknowledgements:** Editorial support was provided by Mark English, PhD of Bellbirdmedical Communications.

**Author Contributions**: K.L. and J.F. designed the study, K.L., X.S., J.M., Z.Z., Q.D., S.P., J.W. and S.T. performed the research, K.L., X.S., G.B. and J.F. analyzed the data and wrote the manuscript.

**Author Information**: K.L. and J.F. are employees of Accutar biotechnology; X.S., J.M., Z.Z., J.W., S.P. and S.T. are employees of Accutar biotechnology(Shanghai); Q.D. and G.B. declare no competing financial interests.

Readers are welcome to comment on the online version of the paper.

Correspondence and requests for materials should be addressed to JF (jiefan@accutarbio.com).

## Tables

## Table 1. Atom Type Mapping table

| Index | Description | Count | Atoms |
|---|---|---|---|
| 0 | Undefined atom type | | |
| 1 | Planar carbon with one single bond and two double bonds | 29 | ALA C; ARG C; ASN C; ASN CG; ASP C; ASP CG; CYS C; GLN C; GLN CD; GLU C; GLU CD; GLY C; HIS C; HIS CG; ILE C; LEU C; LYS C; MET C; PHE C; PHE CG; PRO C; SER C; THR C; TRP C; TRP CG; TYR C; TYR CG; TYR CZ; VAL C; |
| 2 | Tetrahedral carbon with three single bonds | 23 | ALA CA; ARG CA; ASN CA; ASP CA; CYS CA; GLN CA; GLU CA; HIS CA; ILE CA; ILE CB; LEU CA; LEU CG; LYS CA; MET CA; PHE CA; PRO CA; SER CA; THR CA; THR CB; TRP CA; TYR CA; VAL CA; VAL CB; |
| 3 | Carbon with only one single bond | 9 | ALA CB; ILE CD1; ILE CG2; LEU CD1; LEU CD2; MET CE; THR CG2; VAL CG1; VAL CG2; |
| 4 | Backbone nitrogen atom with one double bond and one | 20 | ALA N; ARG N; ARG NE; ASN N; ASP N; CYS N; GLN N; GLU N; GLY N; HIS N; ILE N; LEU N; LYS N; MET N; PHE N; SER N; THR N; TRP |
| | single bond | | N; TYR N; VAL N; |
| 5 | Oxygen atom with one double bond | 26 | ALA O; ARG O; ASN O; ASN OD1; ASP O; ASP OD1; ASP OD2; CYS O; GLN O; GLN OE1; GLU O; GLU OE1; GLU OE2; GLY O; HIS O; ILE O; LEU O; LYS O; MET O; PHE O; PRO O; SER O; THR O; TRP O; TYR O; VAL O; |
| 6 | Carbon with two single bonds | 27 | ARG CB; ARG CD; ARG CG; ASN CB; ASP CB; CYS CB; GLN CB; GLN CG; GLU CB; GLU CG; GLY CA; HIS CB; ILE CG1; LEU CB; LYS CB; LYS CD; LYS CE; LYS CG; MET CB; MET CG; PHE CB; PRO CB; PRO CD; PRO CG; SER CB; TRP CB; TYR CB; |
| 7 | Planar carbon with three double bonds | 3 | ARG CZ; TRP CD2; TRP CE2; |
| 8 | Nitrogen atom with one double bond | 4 | ARG NH1; ARG NH2; ASN ND2; GLN NE2; |
| 9 | Sulfur with one single bond | 1 | CYS SG; |
| 10 | Carbon with two double bonds | 16 | HIS CD2; HIS CE1; PHE CD1; PHE CD2; PHE CE1; PHE CE2; PHE CZ; TRP CD1; TRP CE3; TRP CH2; TRP CZ2; TRP CZ3; TYR CD1; TYR |
| | | | CD2; TYR CE1; TYR CE2; |
| 11 | Nitrogen with two double bonds | 3 | HIS ND1; HIS NE2; TRP NE1; |
| 12 | Nitrogen with one single bond | 1 | LYS NZ; |
| 13 | Sulfur with two single bonds | 1 | MET SD; |
| 14 | Nitrogen with three single bonds | 1 | PRO N; |
| 15 | Oxygen atom with one single bond | 3 | SER OG; THR OG1; TYR OH; |
| 16 | Other carbon atom | | Other C |
| 17 | Other oxygen atom | | Other O |
| 18 | Other nitrogen atom | | Other N |
| 19 | Other sulfide atom | | Other S |
| 20 | Phosphor atom | | P |
| 21 | Halogen atom | | F; CL; BR; I; |
| 22 | Metallic atom | | Mg; Fe; Zn; etc. |

**Figures**

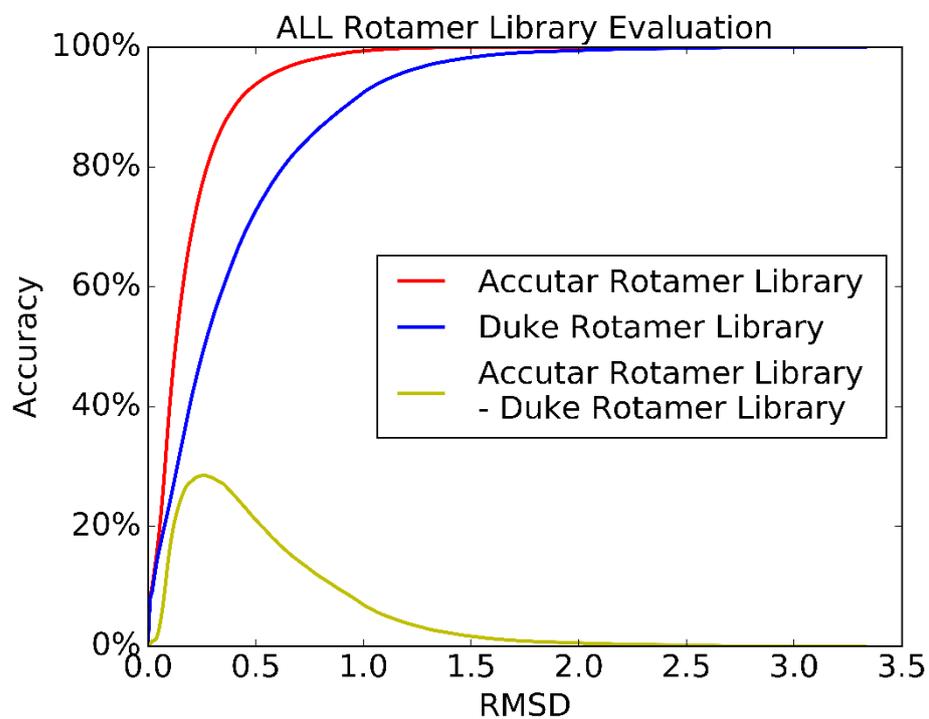

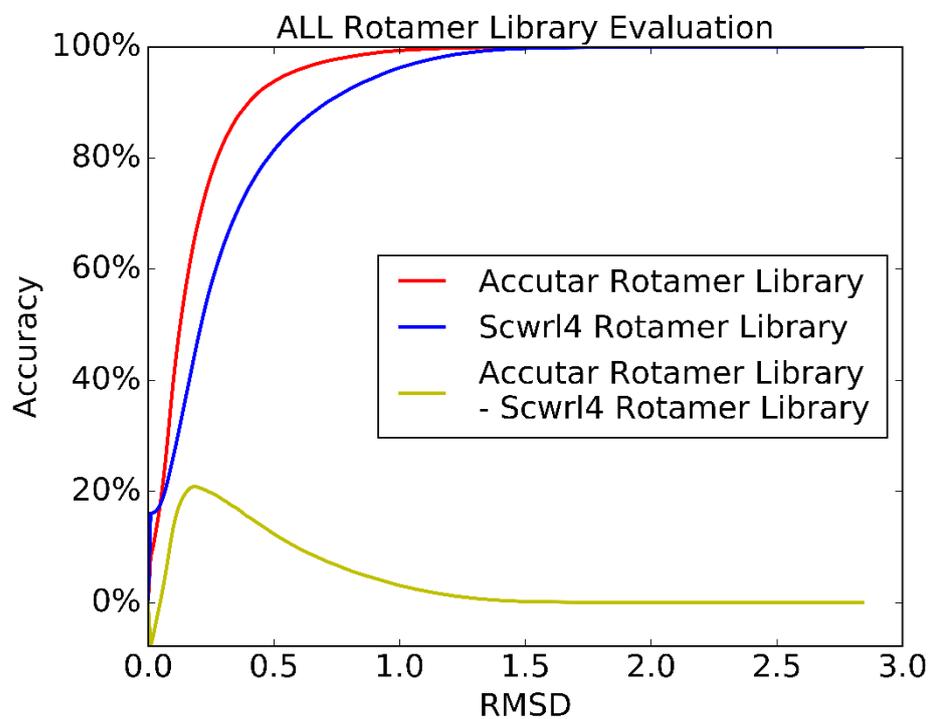

**Fig. 1.**

**a**

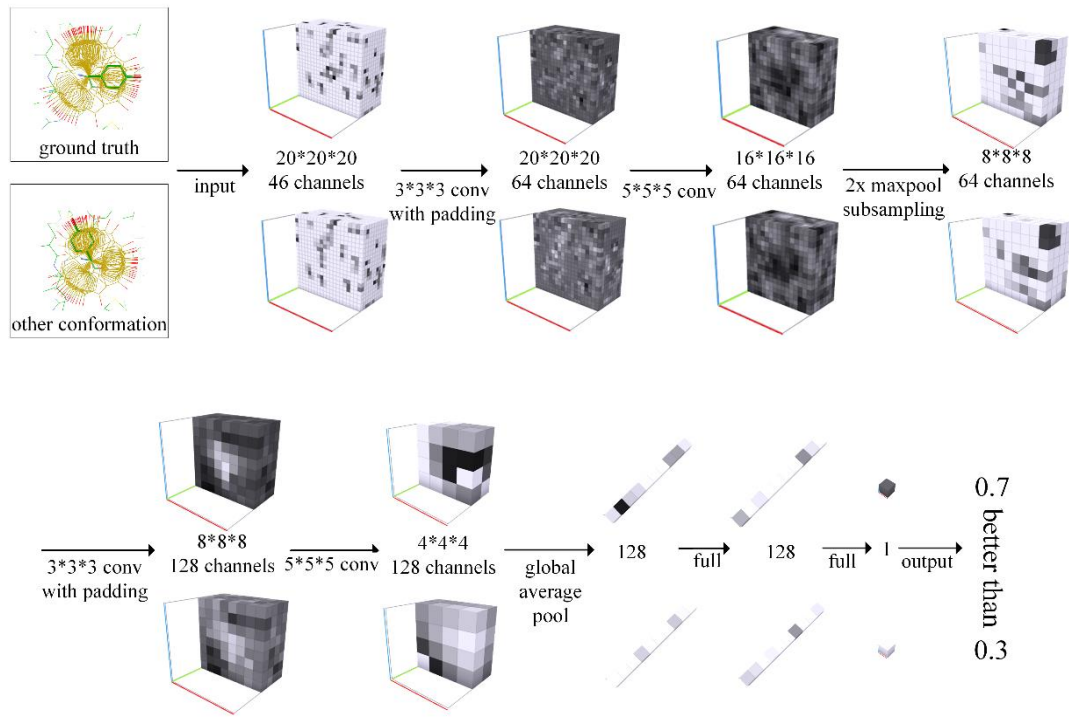

**b**

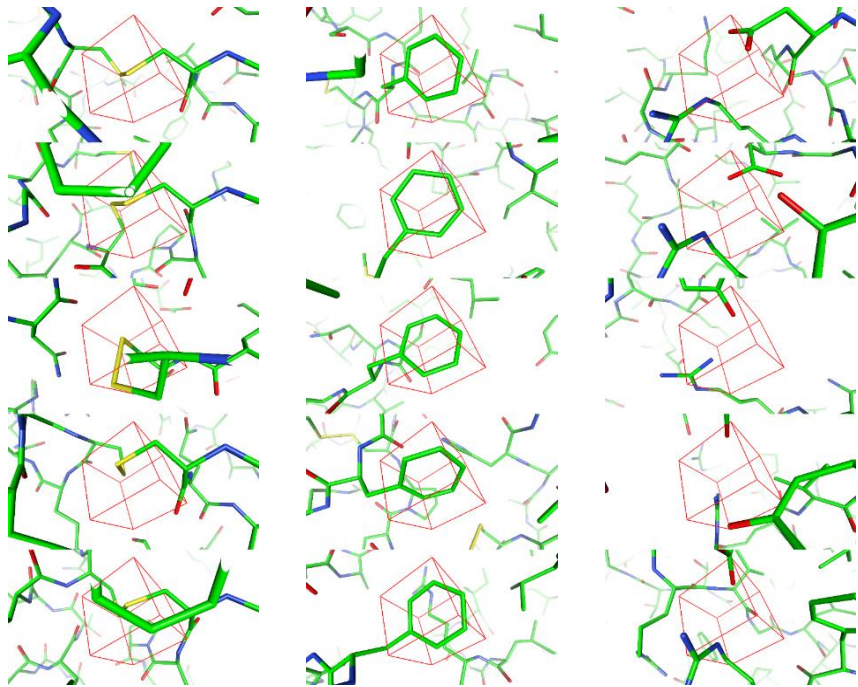

**Fig. 2.**

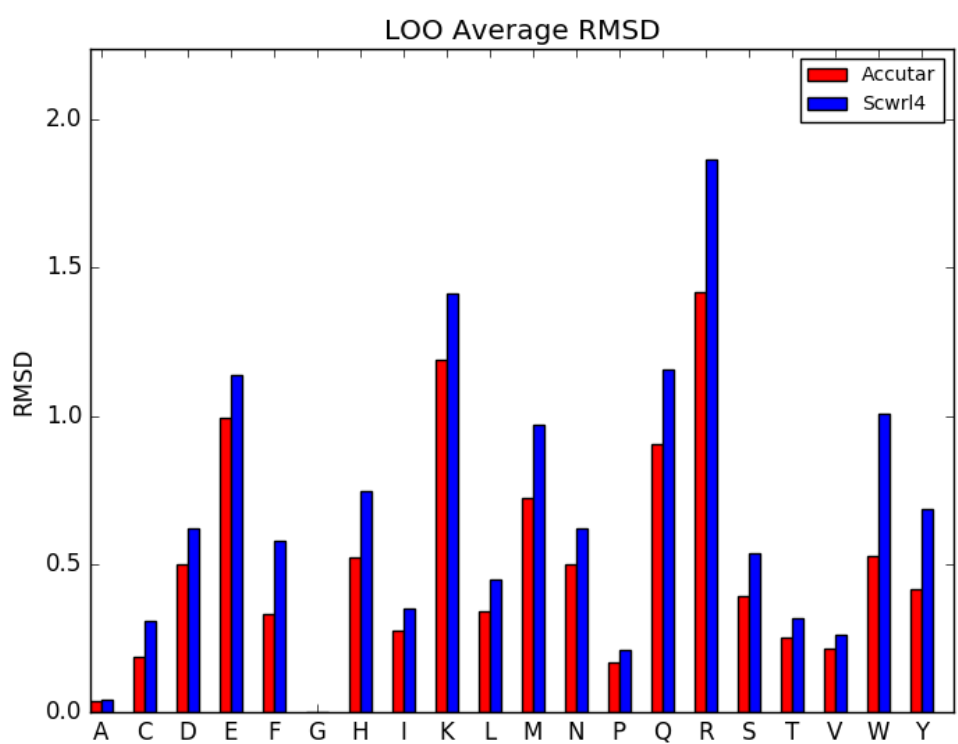

**Fig. 3.**

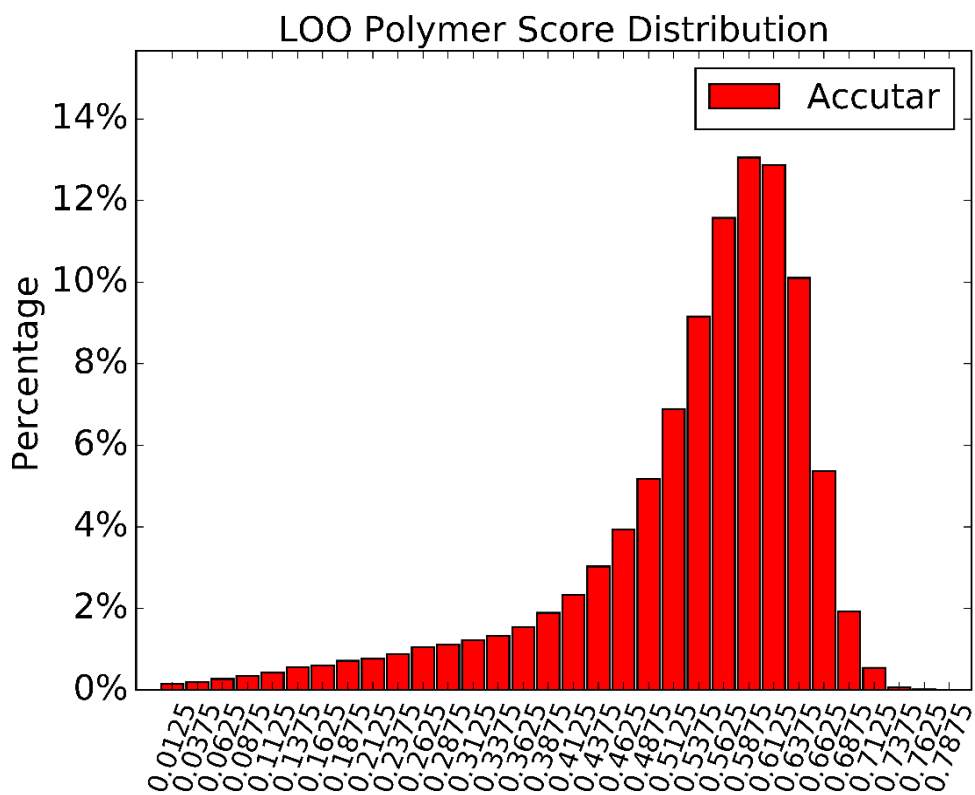

**Fig. 4a**

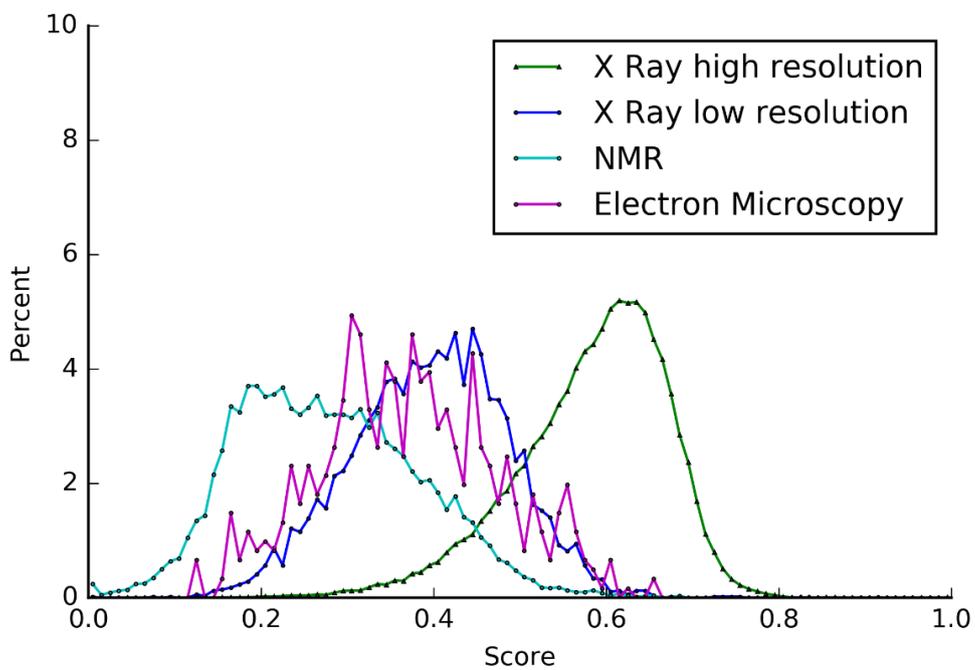

**Fig. 4b**

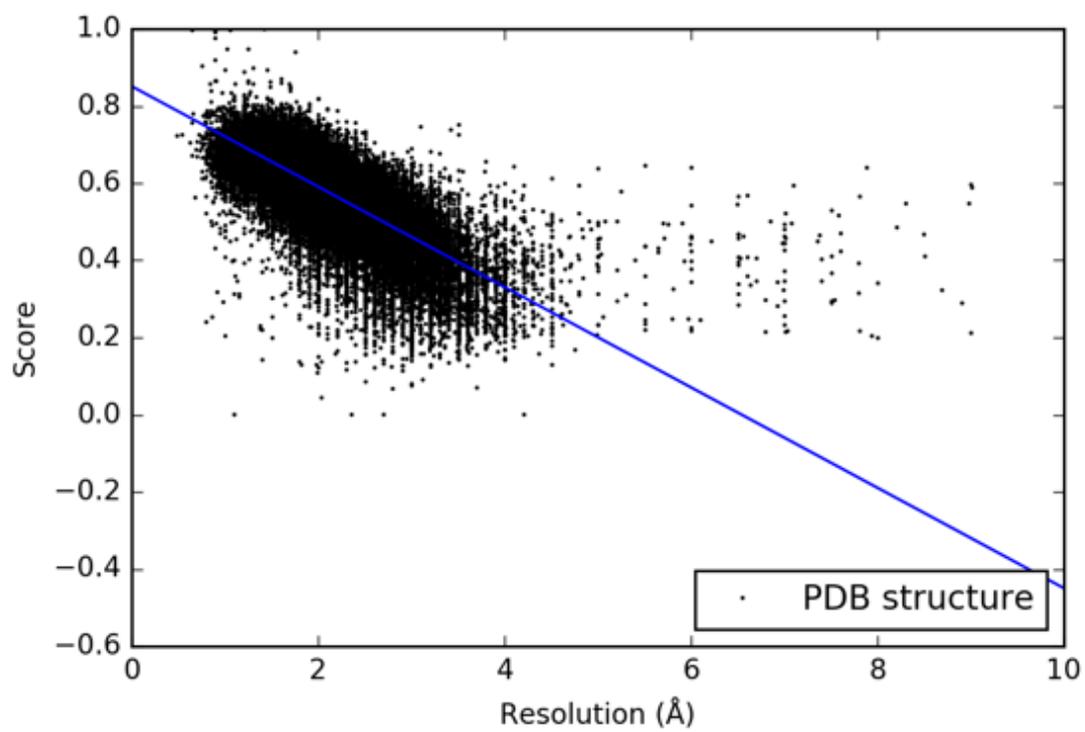

$R^2 = 0.539545$

$y = -0.130070x + 0.851201$

**Fig. 4c**

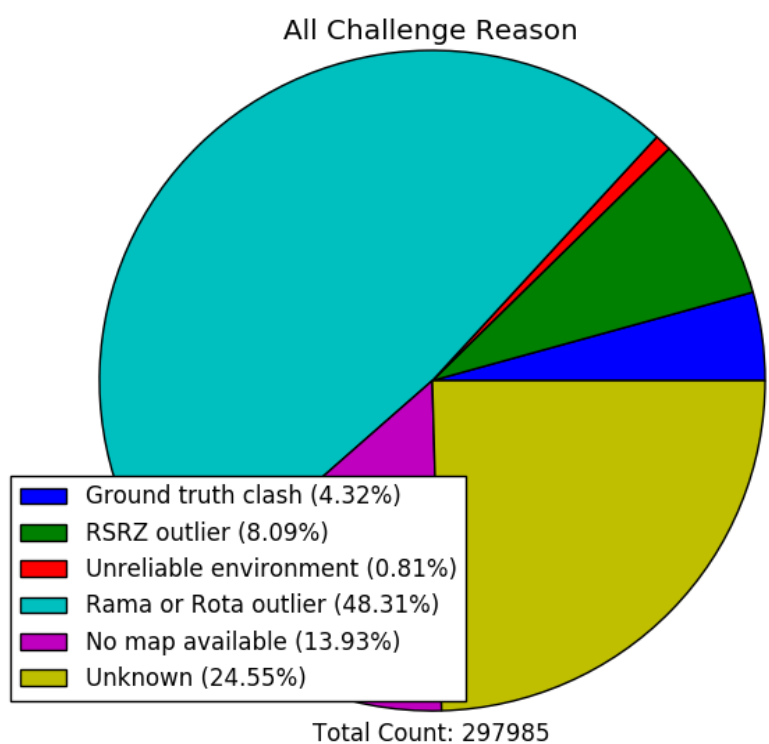

Total Count: 297985

**Fig. 5a**

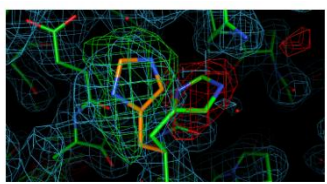

1H49-B58 HIS

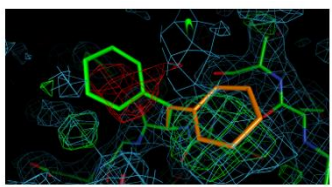

2RCS-L83 PHE

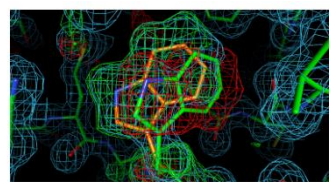

4G95-A113 TRP

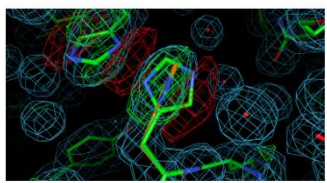

1W27-B486 HIS

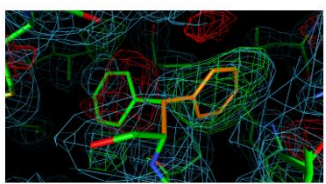

3D3K-D371 PHE

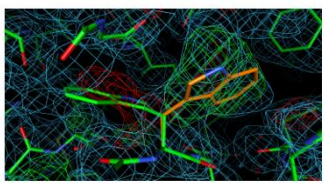

1IVG-B352 TRP

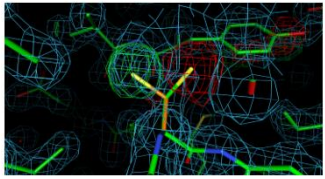
1PS3-A497 CYS

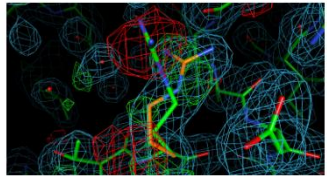
3I8P-A99 ARG

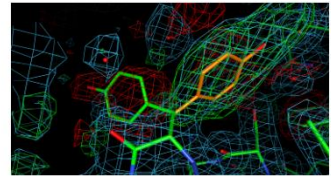
2J8D-L162 TYR

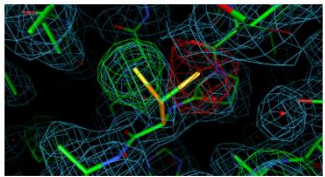
3LF7-A62 CYS

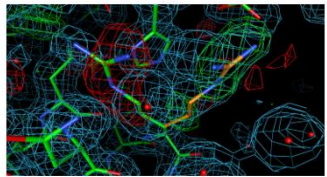
3FPH-B210 ARG

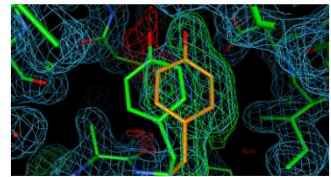
1PQI-A161 TYR

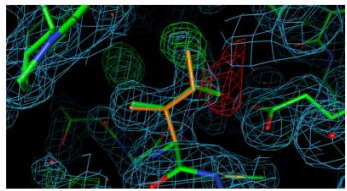
3U8G-B266 ILE

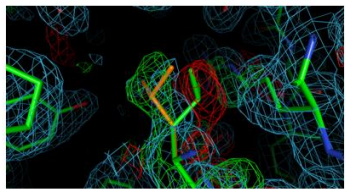
2F26-A217 LEU

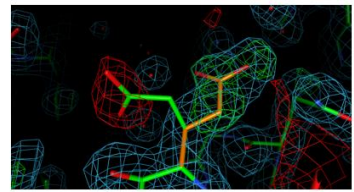
2OVD-A151 GLU

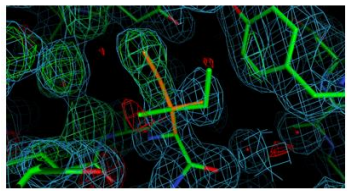
1W7G-H79 ILE

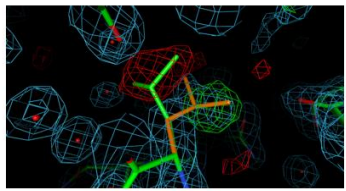
1SKA-A159 LEU

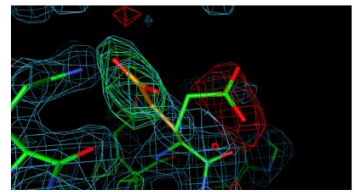
2OVE-A151 GLU

**Fig. 5b**

**Figure legends**

**Fig. 1. | The ACCUTAR side chain rotamer library outperforms current rotamer libraries**

The cumulative distribution function (CDF) plot of the Accutar rotamer library, Duke rotamer library and SCWRL4 rotamer library are shown, with CDF being defined as

$$F_X(x) = P(X_{deviation}(RMSD\ measured\ in\ Å) \leq x)$$

This calculation was based on the assumption that the side chain conformations of all amino acids in a protein have been fully represented by the rotamer nearest to the genuine conformation. Hence, the deviation (measured by RMSD) between the genuine structure and models, represented by different rotamer libraries including the SCWRL4 rotamer library, Duke rotamer library or Accutar rotamer library, could be measured and used to calculate the CDF functions. The CDF functions of the Accutar rotamer library and the rotamer library and their differences are colored by red, green and blue, respectively.

**Fig. 2. | Construction of a Convolutional Neural Network architecture for side chain conformation prediction**

**a,** In our designed deep CNN architecture, data flow is shown from left to right: a rotamer of an amino acid (for example, rotamer # 4 of tyrosine) is represented as a grid of 20*20*20 voxels. In order to convert the discrete amino acid conformer to a dense feature map to CNN, each amino acid rotamer and related environment were encoded by 23 atom type and represented as a smoothly interpolated sphere in the grid using the soft-bin fill algorithm, as shown in **fig. S6**. Atoms of the side chain conformation to be predicted and of its environment were extracted into separated channels to be able to distinguish them. As a result, a total of 46 input channels were used (layer 0). The neural network accepts a voxel grid of the quantized amino acid environment and approximates a piecewise ranking score. The 20*20*20 voxel was fed through a 3*3*3 convolutional layer and a 5*5*5 convolutional layer, with a 2*2*2 max pool subsampling. Then another 3*3*3 and 5*5*5 convolutional layers were applied. Finally, a global average pooling layer was used to aggregate information from the entire grid and several fully connected layers were applied subsequently to project the output to a scalar score. ReLU non-linearity was used throughout the process except the output layer, where a sigmoid non-linearity was used to map the output to probability of range (0, 1).

**b,** Signature input patches indicating chemical properties (disulfide bonds, benzene and electrostatic interaction), which maximally activated a filter in the first convolution layer are shown.

**Fig. 3. | The Accutar platform out performs current methods significantly for side chain conformation prediction**

The prediction accuracy for each amino acid type by different methods were compared by RMSD criteria. All residues from the test set constituting 379 pdbs were used to run a LOO test (see main text). The RMSD for each residue type averaged over all residues are shown in the figure with Accutar method shown in red and SWCRL4 method shown in blue.

**Fig. 4. | The LOO score could be used to judge model quality**

**a**, Pan-PDB side-chain LOO scores could be used to judge model quality.

This figure shows the probability distribution function plot of pan-PDB side-chain LOO scores.

**b**, Probability distribution of LOO scores for all PDBs.

This figure shows the probability distribution of LOO scores categorized by different model types with high resolution (<3A) X-ray model plot in green and low resolution X-ray model plot in blue, EM model plot in cyan and NMR model plot in red.

**c, Scatter plot of X Ray PDB Resolution and Probability distribution of its LOO score**

This figure shows a scatter plot of atomic resolution of X-ray structures and their associated LOO score with an observed $r^2$ of 0.53 and spearman score of 0.75. A linear fitting of the plot is also shown.

**Fig. 5.**

**a, LOO outlier analysis**

This figure shows a pie chart of side chain LOO score outliers of all PDB structures. Statistics based on top amino acid outliers with an associated LOO score beyond 3 standard deviation ranges of the average LOO score of its kind, are shown in the pie chart. The outliers were plotted by following six classes: ground truth clashes (blue), RSRZ outliers(green), unreliable environment (red), Ramachandran/rotamer outlier (cyan) and no map available (purple), unknown (yellow). The calculation of RSRZ, Ramachandran and rotamer outliers uses the same protocol as RCSB X-ray validation process(*27-29*).

**Key:**

**Ground truth clash**: At least one atom in the amino acid has a too close contact with another atom. The close contact may occur inside the amino acid, between this amino and another amino, or between this amino and a hetero.

Both residue and backbone atoms in this amino are checked for clashes.

**RSRZ outlier**: RSRZ is a normalization of real-space R-value (RSR) which measures the quality of fit between the amino acid and the data in real space.

A residue is considered an RSRZ outlier if its RSRZ value is greater than 2.

**Unreliable environment**: an excessive number (>=5) of clashes have been detected near the amino acid (<= 10 A).

**Rama or Rota outlier**: This amino acid is considered a Ramachandran plot outlier (for backbone) or a rotamer outlier (for residue).

The outlier is assessed as with MolProbity.(*29*) This type of outlier indicates the amino acid having unusual torsion angles, not similar to any preferred combinations.

**No map available**: There is no specific errors detected with this amino acid, except the quality of fit between the amino acid and the density map cannot be checked due to the lack of density map data.

**Unknown**: There is no specific errors detected with this amino acid.

**b** The Accutar side chain predictor can predict side chain conformational error of published high resolution crystal structure (examples).

**Supplementary Figures**

**fig. S1. │ The ACCUTAR side chain rotamer library outperforms current rotamer libraries**

CDF plot for each amino acid type in the Accutar rotamer library (shown in red), SCWRL4 rotamer library (shown in blue) and their difference (shown in green) are shown.

**fig. S2. │ CNN ranking model evaluation**

The ranking model used in CNN training algorithm was evaluated by plotting the accuracy at the kth rank. The evaluation metric is similar to precision@k(*30*). For every amino acid in the test set, all rotamers of its kind were retrieved, and then their predicted scores were compared with the predicted score of ground truth. The top K ranked rotamers were used for inspection. In the "ground truth" evaluation scheme, if the ground truth occurs in the top K ranked rotamers, the scoring for this amino acid was considered correct. In the "similar rotamer" evaluation scheme, if any rotamer with an RMSD with ground truth less than a predefined small value is within the top K ranked rotamers, the scoring for this amino acid was considered correct. The accuracy for the entire test set was then defined as the average correctness rate for each amino acid type.

**fig. S3. │ ACCUTAR out performs the SCWRL4 method by RMSD criteria**

CDF function for each amino acid prediction accuracy rate with respect RMSD were plotted with the Accutar model shown in red, the SCWRL4 model in blue and their difference shown in yellow. CDF function plots for the Accutar rotamer library and Duke rotamer library is also shown in the graph for comparison.

**fig. S4. | Histogram of Probability Score for all PDBs**

This figure is related to Figure 4B. The probability distribution function of the LOO scores for different model types were individually plotted by histogram.

**fig. S5. | LOO outlier analysis**

This figure contains pie charts of LOO outliers for each amino acid related to **Figure. 5a**. The pie chart of the LOO outliers for each amino acid type were created using same color label as in **Figure 5a.**

**fig. S6. | The soft-bin grid fill algorithm**

**fig. S7. | The perturbation algorithm**

**fig. S8. | The fine tuning algorithm**

**Supplementary Tables**

**Supplementary Table 1 ｜ Chi angles of each rotamer for each amino acid in the Accutar rotamer library**

This table lists the content Accutar rotamer library. Each line shows a single rotamer. A rotamer is given as its dihedral angles in radians.

**Supplementary Table 2 ｜ LOO outlier list**

This table lists all LOO outliers found in all PDB structures. An amino acid is identified as an outlier if its associated LOO score is beyond 3 standard deviation ranges of the average LOO score of its kind.

**Key:**

**PDBID**: RCSB PDB ID.

**Method**: The reported experimental method for obtaining the PDB structure.

**ResID**: The residue ID of this amino acid in PDB file.

**Sigma**: How many standard deviations the LOO score of the amino acid is beyond the average LOO score of its amino kind. The larger the sigma, the worse the LOO score.

**Reason**: The reason label of this amino acid.